\documentclass[twocolumn,showpacs,preprintnumbers,amsmath,amssymb]{revtex4}

\usepackage{graphicx}
\usepackage{dcolumn}
\usepackage{bm}

\usepackage{amssymb}

\begin{document}

\title{Quantum cosmology,  minimal length and  holography}

\author{S. Jalalzadeh}
\email{s-jalalzadeh@sbu.ac.ir}
 \affiliation{Department of Physics, Shahid Beheshti University, G. C., Evin, 19839 Tehran, Iran.}
\author{S. M. M. Rasouli$^{1,2}$}

\email{mrasouli@ubi.pt}
\author{P. V. Moniz$^{1,2}$}
\email{pmoniz@ubi.pt}
\affiliation{$^1$Departamento de F\'{i}sica, Universidade da Beira Interior, Rua Marqu\^{e}s d'Avila
e Bolama, 6200 Covilh\~{a}, Portugal.}

\affiliation{$^2$Centro de Matem\'{a}tica e Aplica\c{c}\~{o}es (CMA - UBI),
Universidade da Beira Interior, Rua Marqu\^{e}s d'Avila
e Bolama, 6200 Covilh\~{a}, Portugal.}

\date{\today}

\begin{abstract}
We study the effects of a generalized uncertainty principle on the classical
and quantum cosmology of a closed Friedmann Universe
whose matter content is either a dust or a radiation fluid.
More concretely, assuming the existence of a minimal length,
we show that the entropy will constitute a
Dirac observable.
In addition, 't Hooft conjecture on the cosmological
holographic principle is also investigated .
We describe how this holographic principle is satisfied for large values of a
 quantum number, $n$. This occurs when the entropy is computed in terms
of the minimal area.
\end{abstract}

\maketitle
\section{Introduction}
\label{int}

It has been pointed that our existing
theories will break down when applied to small distances or very high energies.
In particular, the geometrical continuum beyond
a certain limit will no longer be valid.
This suggested the development of a scenario
based on indivisible units of length. In recent years,
the concept of minimal length has been described through algebraic
methods, by means of a generalized uncertainty principle, which could be induced by gravitational
effects, first proposed by Mead \cite{M1}. Moreover, a generalized uncertainty
principle (GUP), as presented in theories such as string
theory and doubly special relativity, conveys the prediction of a minimal measurable
length \cite{M2}. A similar feature  appears in the
polymer quantization in terms of a mass scale \cite{M3}. The concept of a subsequent minimal
area can, thus, be considered. This then raises the question of whether
it can be used in discussions about entropy, namely, involving the holographic principle.

The holographic principle in quantum gravity was first suggested by 't Hooft
\cite{Hooft} and later extended to string theory by Susskind \cite{Sus}.
The most radical part of this principle proposes that the
degrees of freedom of a spatial region reside not in the bulk but in the boundary.
Furthermore, the number of boundary degrees of freedom per
Planck area should not be larger than one. In this context, it is worth
to remind the reader the general assumption that the Bekenstein-Hawking area
law applies universally to all cosmic or black hole horizons.
On the other hand, it has been shown recently, (in Ref.~\cite{L13}) that there is a derivation for the
holographic/conformal-anomaly Friedmann equation.
This is of interest because that derivation was obtained by assuming that the effect of
 GUP on the entropy from the apparent horizon and admits a
constraint which relates the anomaly coefficient
and the GUP parameter.

In this paper we investigate the quantum cosmology of a
closed~Friedmann--Lema\^{\i}tre--Robertson--Walker~(FLRW) Universe, filled
with either  radiation
 or  dust. Our aim in this scenario, by assuming a minimal length, is to determine
whether a corresponding minimal area will be of relevance in discussing the holographic principle.
The paper is organized as follows. In Section \ref{CGUP} we present the classical setting
in the presence of a GUP. Section \ref{QCML} provides the quantum cosmological description
of our model. Section \ref{HP} conveys how our model can be used to discuss holographic features.
In Section \ref{con}, we summarize our results.

\section{classical model with GUP}
\label{CGUP}

Let us start with the line element of the closed homogeneous and isotropic FLRW geometry
\begin{eqnarray}\label{1}
ds^2=-N^2(\eta)d\eta^2+a^2(\eta)d\Omega^2_{(3)},
\end{eqnarray}
where $N(\eta)$ is the lapse function, $a(\eta)$ is the scale factor and
$d\Omega^2_{(3)}$ is the standard line element for a unit three-sphere.
The action functional corresponding to the line element (\ref{1}) displays in the gravitational
and matter sectors (with the latter as perfect fluid)~\cite{Hawking}
\begin{eqnarray}\label{2}
\begin{array}{cc}
{\cal S}=\frac{M_{\text{Pl}}^2}{2}\int_{\cal M}\sqrt{-g}Rd^4x\\
\\
+M_{\text{Pl}}^2\int_{\partial{\cal
M}}\sqrt{g^{(3)}}Kd^3x
-\int_{\cal M}\sqrt{-g}\rho d^4x\\
\\=6\pi^2M_{\text{Pl}}^2\int\left(-\frac{a\dot{a}^2}{N}+Na\right)d\eta-2\pi^2\int
Na^3\rho d\eta,
\end{array}
\end{eqnarray}
 in which $M_{\text{Pl}}^2=\frac{1}{8\pi G}$ (where $G$ is the Newton gravitational constant)
 is the reduced Planck's mass squared in
 natural units, ${\cal M}=I\times S^3$ is the spacetime manifold, $\partial{\cal
 M}=S^3$, $R$ is the Ricci scalar associated to the metric $g_{\mu\nu}$ whose
 determinant has been denoted by $g$, $\rho$ is the energy density, $K$ is the trace of the
 extrinsic curvature of the spacetime boundary
 and an overdot denotes differentiation with respect to $\eta$.

For a radiation fluid ($p_\gamma=\frac{1}{3}\rho_\gamma$ where
the $\rho_\gamma$ and $p_\gamma$ are the energy density and
pressure associated to the radiation fluid, respectively),
redefining the scale factor and lapse function as
 \begin{eqnarray}\label{6}
 \begin{cases}
 a(\eta)=x(\eta)+\frac{M}{12\pi^2M_{\text{Pl}}^2}:=x-x_{0},\\
 N(\eta)=12\pi^2M_{\text{Pl}}a(\eta)\tilde{N},
 \end{cases}
 \end{eqnarray}
 the total Lagrangian,
 if we further add a dust fluid, non-interacting with the radiation,
 will be
\begin{eqnarray}\label{7}
 {\mathcal L}=-\frac{1}{2\tilde{N}}M_{\text{Pl}}\dot{x}^2+\frac{\tilde{N}}{2}M_{\text{Pl}}\omega^2x^2-{\mathcal
 E}\tilde{N},
 \end{eqnarray}
 where we
have employed
 \begin{eqnarray}\label{8}
 \begin{cases}
 {\mathcal E}=\frac{M^2}{2M_{\text{Pl}}}+12\pi^2{\mathcal N}_\gamma M_{\text{Pl}},\\

 \omega=12\pi^2M_{\text{Pl}}.
 \end{cases}
 \end{eqnarray}
 Moreover, we introduce $M$ and ${\mathcal N}_\gamma$ as
   \begin{eqnarray}
 \begin{cases}
 M=\int_{\partial {\cal M}}\sqrt{g^{(3)}}\rho_{0m}a_0^3d^3x,\\
 {\mathcal N}_\gamma=\int_{\partial {\cal M}}\sqrt{g^{(3)}}\rho_{0\gamma}a_0^4d^3x.
 \end{cases}
 \end{eqnarray}
In the above definition, $M$ denotes the total mass of
the dust matter.
In addition, ${\mathcal N}_\gamma$ could be related to the total entropy of radiation
as follows: the energy density $\rho_\gamma$,
the number density $n_\gamma$, the entropy density $s_\gamma$
and the scale factor are related to temperature, $T$, via $\rho_\gamma=\frac{\pi^2}{30}gT^4$,
$n_\gamma=\frac{\zeta(3)}{\pi^2}gT^3$, $s_\gamma=\frac{4}{3}\frac{\rho_\gamma}{T}$
and $a(\eta)\sim\frac{1}{T}$~\cite{Mukhanov}. Consequently, we find
\begin{eqnarray}\label{RR1}
{\mathcal
N}_\gamma=\left(\frac{5\times3^5}{2^8\pi^4 g}\right)^{1/3}\left(S^{(\gamma)}\right)^{4/3},
\end{eqnarray}
 where $S^{(\gamma)}$ denotes the
 entropy~\cite{footnote} corresponding to the radiation fluid.

The momenta conjugate to $x$ and the primary
constraint, which are necessary to construct the Hamiltonian of the model, are given by
 \begin{eqnarray}\label{9}
 \begin{cases}
 \Pi_x=\frac{\partial{\mathcal L}}{\partial\dot{x}}=-\frac{\tilde{N}}{M_{\text{Pl}}}\dot{x},\\

 \Pi_{\tilde{N}}=\frac{\partial{\mathcal L}}{\partial\dot{\tilde{N}}}=0.
 \end{cases}
 \end{eqnarray}
Therefore, the Hamiltonian corresponding to
 (\ref{7}) becomes
 \begin{eqnarray}\label{10}
 {\mathcal H}=-\tilde{N}\left[\frac{1}{2M_{\text{Pl}}}\Pi^2_x+\frac{1}{2}M_{\text{Pl}}\omega^2x^2-{\mathcal
 E}\right].
 \end{eqnarray}
From (\ref{9}) we see that the momentum conjugate to $\tilde{N}$ vanishes, i.e.,
the Lagrangian of the system is singular.
Thus, we have to add it to the Hamiltonian~(\ref{10}) and construct the total
Hamiltonian as
\begin{eqnarray}\label{11}
{\mathcal H}_{\text{T}}=-\tilde{N}\left[\frac{1}{2M_{\text{Pl}}}\Pi^2_x+\frac{1}{2}M_{\text{Pl}}\omega^2x^2-{\mathcal
 E}\right]+\lambda\Pi_{\tilde{N}},
\end{eqnarray}
where $\lambda$ is a Lagrange multiplier.

During the evolution of the system, the primary constraint should  hold; namely, we should have
\begin{eqnarray}\label{12}
\dot\Pi_{\tilde{N}}=\{\Pi_{\tilde{N}},{\mathcal H}_{\text{T}}\}\approx0,
\end{eqnarray}
which leads to the secondary (Hamiltonian) constraint as
\begin{eqnarray}\label{13}
H:= \frac{1}{2M_{\text{Pl}}}\Pi^2_x+\frac{1}{2}M_{\text{Pl}}\omega^2x^2-{\mathcal
 E}\approx0.
\end{eqnarray}
We should note that a gauge-fixing condition is required for the constraint (\ref{13}), in which
 $\tilde{N}=const.$ can be a possibility. Thus, by choosing the gauge as $\tilde{N}=1/\omega$, and reminding that
the canonical variables satisfy the Poisson algebra $\{x,\Pi_x\}=1$, we get the following Hamilton equations of motion
\begin{eqnarray}\label{14}
\begin{cases}
\dot{x}=-\frac{1}{\omega M_{\text{Pl}}}\Pi_x,\\

\dot{\Pi}_x=\omega M_{\text{Pl}}x.
\end{cases}
\end{eqnarray}
Employing  the Hamiltonian constraint (\ref{13}), it
 easily leads us to the well known solution for a closed Universe as
\begin{eqnarray}\label{15}
\begin{cases}
a(\eta)=\frac{a_{\text{Max}}}{1+\sec\phi}\left[1-\sec\phi\cos(\eta+\phi)\right],\\
a_{\text{Max}}:=\frac{M}{12\pi^2M_{\text{Pl}}^2}+\left(\frac{2{\mathcal E}}
{M_{\text{Pl}}\omega^2}\right)^{\frac{1}{2}},\\
\cos\phi:=\frac{M}{\sqrt{2{\mathcal E}M_{\text{Pl}}}},
\end{cases}
\end{eqnarray}
where $a_{\text{Max}}$ represents the maximum radius of
the closed Universe and we have assumed that
the initial singularity occurs at $\eta=0$.

 Let us now investigate the effects, at a classical level,
 of a deformed Poisson algebra in the presence of a minimal
length. We write \cite{deformed}
\begin{eqnarray}\label{16}
\begin{cases}
\{x,x\}=\{\Pi_x,\Pi_x\}=0,\\
\\
\{x,\Pi_x\}=1+\alpha^2L_{\text{Pl}}^2\Pi_x^2,
\end{cases}
\end{eqnarray}
where $\alpha$ is a dimensionless constant  and $L_{\text{Pl}}$ denotes
Planck's length in the natural units. It is normally assumed that $\alpha$
is to be of order unity.
In this case, the deformation will contribute only to the Planck regime of the
Universe, and for this reason, the quantum cosmology of
the  model will be studied in the next section.
A physical length of the order $\alpha L_{\text{Pl}}$ is yet
 unobserved so it cannot exceed the electroweak scale \cite{elect},
which implies $\frac{1}{\sqrt{8\pi}}\leqslant\alpha\leqslant10^{17}$.
As a consequence of the above deformation with
Hamiltonian (\ref{10}), the Hamilton equations become
\begin{eqnarray}\label{17}
\begin{cases}
\dot{x}=-\frac{\tilde{N}}{M_{\text{Pl}}}(1+\alpha^2L_{\text{Pl}}^2\Pi_x^2)\Pi_x,\\
\\
\dot{\Pi}_x=\tilde{N}\omega^2M_{\text{Pl}}(1+\alpha^2L_{\text{Pl}}^2\Pi^2_x)x.
\end{cases}
\end{eqnarray}
To solve these equations, we relate the $\Pi_x$ to the new variable $y$ as
\begin{eqnarray}\label{18}
\Pi_x=\frac{1}{\alpha L_{\text{Pl}}}\tan(\alpha L_{\text{Pl}}y),
\end{eqnarray}
which, using the Hamiltonian constraint (\ref{13}) and the gauge $\tilde{N}=1/\omega$, gives
\begin{eqnarray}\label{19}
\begin{cases}
a(\eta)=\frac{a_{\text{Max}}}{B}\left(1-\frac{A\cos({\Omega\eta+\phi)}}
{\sqrt{1+2{\mathcal E}\alpha^2L_{\text{Pl}}\cos^2(\Omega\eta+\phi)}}\right),\\
B:=1+\Omega^{-1}\sqrt{{1+2\alpha^2L_{\text{Pl}}{\mathcal E}\cos^2\phi}}\sec\phi,\\
A:=\sec\phi{\sqrt{1+2\alpha^2L_{\text{Pl}}{\mathcal E}\cos^2\phi}},\\
\Omega:=\sqrt{1+2{\mathcal E}\alpha^2L_{\text{Pl}}},\\
\cos\phi:=\frac{M}{\sqrt{2{\mathcal E}M_{\text{Pl}}}}\left(1+24\pi^2\alpha^2{\mathcal
N}_\gamma\right)^{-\frac{1}{2}},
\end{cases}
\end{eqnarray}
and $a_{\text{Max}}$ is similar to the non-deformed case defined in (\ref{15}).
If we take the limit $\alpha\rightarrow0$, we find solution  (\ref{15})
which shows that the canonical behavior is recovered in this limit.
 We immediately obtain from (\ref{19}) that the Universe reaches its maximum
 radius at $\eta=\frac{\pi-\phi}{\Omega}$, and it terminates in the big-crunch
 singularity at $\eta=\frac{2(\pi-\phi)}{\Omega}$.

\section{Quantum cosmology with minimal length}
\label{QCML}

At the quantum level, the deformed Poisson
algebra (\ref{16}) is replaced by the following commutation
relation between the phase space variables of the minisuperspace
\begin{eqnarray}\label{20}
[x,\Pi_x]=i(1+\alpha^2L_{\text{Pl}}^2\Pi^2_x).
\end{eqnarray}
Commutation relation (\ref{20}) provides the minimal length uncertainty relation (MLUR) \cite{MLUR}
\begin{eqnarray}\label{21}
\Delta x\geq\frac{1}{2}\left(\frac{1}{\Delta\Pi_x}+\alpha^2L_{\text{Pl}}^2\Delta\Pi_x\right).
\end{eqnarray}
This MLUR implies the existence of a minimal length
\begin{eqnarray}\label{22}
\Delta x_{\text{min}}=\alpha L_{\text{Pl}},
\end{eqnarray}
which indicates it is impossible to consider any physical state as the eigenstate
of the position operator \cite{position}. Consequently, working with
the position representation $|x\rangle$ is impossible. In the
presence of  GUP, in order
to recover the information on the spatial distribution of the
quantum system, we would introduce a quasiposition representation, which consists of the projection
of the states onto a set of maximally localized
states $|\psi^{ml}_\xi\rangle$ \cite{GUP}. These states are the proper physical states around
a position $\xi$ with property $\langle\psi^{ml}_\xi|x|\psi^{ml}_\xi\rangle=\xi$
and $(\Delta x)_{|\psi^{ml}_\xi\rangle}=\Delta x_{\text{min}}$.
Thus, the quasiposition wave function will be
\begin{eqnarray}\label{ZZ5}
\psi(\xi)=\int_{-\infty}^\infty\frac{d\Pi_x e^{\frac{i\xi}{\alpha
L_{\text{Pl}}}\tan^{-1}(\alpha L_{\text{Pl}}\Pi_x)}}{(1+\alpha^2L^2_{\text{Pl}}\Pi_x^2)^{\frac{3}{2}}}\psi(\Pi_x),
\end{eqnarray}
 which is a generalization of the Fourier transformation.
 The Hilbert space, in the quasiposition representation,
 is the space of  functions with the usual $\mathbb{L}^2$ norm.
On a dense domain of the Hilbert
space, the position and momentum operators obeying relation (\ref{20}) could
be represented in momentum space as \cite{Rep}
\begin{eqnarray}\label{23}
\begin{cases}
\Pi_x=\Pi_x,\\
x=i(1+\alpha^2L_{\text{Pl}}^2\Pi_x^2)\frac{d}{d\Pi_x}.
\end{cases}
\end{eqnarray}
Hence, the inner product between two arbitrary states on a dense domain will
be
\begin{eqnarray}\label{24}
\langle\varphi|\psi\rangle=\int_{-\infty}^{+\infty}\frac{d\Pi_x}
{1+\alpha^2L_{\text{Pl}}^2\Pi^2_x}\varphi^*(\Pi_x)\psi(\Pi_x).
\end{eqnarray}
Therefore, the modified Wheeler-DeWitt (WDW) equation in the the presence of MLUR (\ref{21}) is given by
\begin{eqnarray}\label{25}
\begin{array}{cc}
\left[-M_{\text{Pl}}\omega^2\left((1+\alpha^2L_{\text{Pl}}^2\Pi_x^2)
\frac{d}{d\Pi_x}\right)^2+\frac{1}{2M_{\text{Pl}}}\Pi_x^2\right]\psi\\
={\mathcal E}\psi.
\end{array}
\end{eqnarray}
We now proceed using the  transformation
 (\ref{18}). Hence, the above WDW equation will be changed
to the trigonometric P\"ochl-Teller (TPT) equation
\begin{eqnarray}\label{26}
\frac{d^2\psi(z)}{dz^2}+\left(\epsilon-\frac{V}{\cos^2(z)}\right)\psi(z)=0,
\end{eqnarray}
where $z=\alpha L_{\text{Pl}}y$ and
\begin{eqnarray}\label{26a}
\begin{cases}
\epsilon=\frac{1}{(12\pi^2\alpha)^2}\left(\frac{2{\mathcal E}}
{M_{\text{Pl}}}+\frac{1}{\alpha^2}\right),\\
V=\frac{1}{(12\pi^2\alpha^2)^2}.
\end{cases}
\end{eqnarray}
The normalized eigenfunctions of the TPT equation are given by \cite{TPT}
\begin{eqnarray}\label{27}
\begin{array}{cc}
\psi_n(z)=
2^\nu\Gamma(\nu)\sqrt{\frac{\nu!(n+\nu)\alpha L_{\text{Pl}}}{2\pi\Gamma(n+2\nu)}}
\cos^\nu(z)C_n^\nu(\sin(z)),
\end{array}
\end{eqnarray}
where $n$ is an integer, $C^\nu_n$ is the Gegenbauer polynomial and
\begin{eqnarray}\label{28}
\nu:=\frac{1}{2}\left(1+\sqrt{1+\frac{1}{36\pi^4\alpha^4}}\right).
\end{eqnarray}
Moreover, the corresponding eigenvalue of the WDW equation is given by
\begin{eqnarray}\label{30m}
\begin{array}{cc}
{\mathcal E}_n=12\pi^2M_{\text{Pl}}\{(n+\frac{1}{2})\sqrt{1+36\pi^4\alpha^4}\\
+6\pi^2\alpha^2(n^2+n+\frac{1}{2})\}=72\pi^4\alpha^2M_{\text{Pl}}(n^2+2n\nu+\nu).
\end{array}
\end{eqnarray}

Let us now obtain the Dirac observables of the model. According to
Dirac~\cite{Dirac}, the observables of a theory are those quantities which have vanishing
 commutators with the constraints of theory.
In order to retrieve them, we start by finding
the symmetries  of the WDW equation in the form given by (\ref{26}). Considering
the infinite number of bound states of Eq.~(\ref{26}), the
underlying Lie algebra could be expected as its spectrum generating algebra.
The lowering and raising operators for the WDW equation in (\ref{26}) can be
built using a factorization type method~\cite{P.Moniz}.
To do this, let us start with the WDW Eq. (\ref{26}), and rewrite
it as a bound state stationary Schr\"odinger equation
\begin{eqnarray}\label{R1}
\begin{cases}
h\psi_n=\epsilon_n\psi_n,\\
h:=-\frac{d^2}{dz^2}+U(z),
\end{cases}
\end{eqnarray}
where $U(z):=V/\cos^2(z)=\nu(\nu-1)/\cos^2(z)$. Introducing the following
first order differential operators~\cite{P.Moniz}
\begin{eqnarray}\label{R2}
a^{\pm}_\nu:=\mp\frac{d}{dz}+W(z;\nu),
\end{eqnarray}
where $W(z;\nu)=\nu\tan(z)$ is the superpotential, we obtain the first supersymmetric
partner Hamiltonian
\begin{eqnarray}\label{R3}
h_+:=a^+_\nu a^-_\nu=h-\epsilon_0,
\end{eqnarray}
where $\epsilon_0=\nu^2$ is the ground state energy eigenvalue. The second
supersymmetric partner Hamiltonian is given by
\begin{eqnarray}\label{R4}
h_-:=a^-_\nu a^+_\nu.
\end{eqnarray}
The Hamiltonians $h_{\pm}$ have the same energy spectrum except the ground
state of $h_+$
\begin{eqnarray}\label{R5}
\begin{cases}
h_+\psi_\nu^n=a^+_\nu a^-_\nu\psi_\nu^n=(\epsilon_n-\nu^2)\psi_\nu^n,\\
h_-\psi_\nu^n=a^-_\nu a^+_\nu\psi_{\nu-1}^n=(\epsilon_n-\nu^2)\psi_{\nu-1}^n.
\end{cases}
\end{eqnarray}
We see that changing the order of operators $a^+_\nu$ and $a^-_\nu$ simply
leads to the shift of the value of $\nu$. This symmetry is called shape-invariance
symmetry~\cite{P.Moniz}. Using shape-invariance symmetry, we can show that $\psi_\nu^n$
and $\psi_{\nu-1}^n$ are proportional to $a^+_\nu\psi_{\nu-1}^n$ and $a^-_\nu\psi_\nu^n$,
respectively. The shape-invariance condition (\ref{R5}) can be rewritten
as
\begin{eqnarray}\label{R6}
a^-_\nu a^+_\nu=a^+_{\nu+1}a^-_{\nu+1}+R(\nu),
\end{eqnarray}
where $R(\nu)=2\nu+1$ is independent of $z$. According to~\cite{Ladder},
we assume that replacing $\nu$ with $\nu+1$ in a given operator, say ${\mathcal
O}_\nu$, can be achieved
with a similarity transformation $T_\nu$:
\begin{eqnarray}\label{R7}
\begin{cases}
T_\nu{\cal O}_\nu(z) T^{-1}_\nu={\cal O}_{\nu+1}(z),\\
T_\nu:=\exp(\frac{\partial}{\partial\nu}).
\end{cases}
\end{eqnarray}
The shape-invariant potentials are easy to deal with, if lowering and raising
operators are employed, developed originally for the harmonic
oscillator. However, as the commutator $[a^-_\nu,a^+_\nu]$
does not yield a constant value, namely,
\begin{eqnarray}\label{RR8}
[a^-_\nu,a^+_\nu]=\frac{2\nu}{\cos^2(z)},
\end{eqnarray}
 the choice of $a^{\pm}_\nu$ does not work.
To establish a suitable algebraic structure, we introduce the following operators
\cite{Ladder}
\begin{eqnarray}\label{30a}
\begin{cases}
A:=T^\dagger a^-_\nu,\\
A^\dagger:=a^+_\nu T,
\end{cases}
\end{eqnarray}
which lead
\begin{eqnarray}\label{R8}
[A,A^\dagger]=1-2\nu.
\end{eqnarray}

The action of these factor operators on normalized eigenfunctions will
be
\begin{eqnarray}\label{30b}
\begin{cases}
A|\nu,n\rangle=\sqrt{n(2\nu+2+n)}|\nu+2,n-1\rangle,\\
A^\dagger|\nu,n\rangle=\sqrt{(n+1)(2\nu-1+n)}|\nu-2,n+1\rangle.
\end{cases}
\end{eqnarray}
It can be verified that these operators together with $\tilde A|\nu,n\rangle=(1/2-\nu)|\nu,n\rangle$,
obey the $su(2)$ Lie algebra
\begin{eqnarray}\label{30c}
[\tilde A,A]=-A,[\tilde A,A^\dagger]=A^\dagger, [A,A^\dagger]=-2\tilde A.
\end{eqnarray}
Also, based on recursion relations of Gegenbauer polynomials, we can
introduce the following three operators \cite{Factorization}
associated with the dynamical group $su(1,1)$
of the WDW equation (\ref{26})
\begin{eqnarray}\label{30}
\begin{cases}
J_+:=\left[-\cos (z)\frac{d}{dz}+\sin(z)(\hat{N}+\nu)\right]\frac{1+\hat{N}+\nu}{\sqrt{(\hat{N}+\nu)(\hat{N}+2\nu)}},\\
J_-:=\left[\cos(z)\frac{d}{dz}+\sin(z)(\hat{N}+\nu)\right]\sqrt{\frac{\hat{N}+\nu-1}{\hat{N}+2\nu-1}},\\
J_0:=\hat{N}+\frac{\nu+1}{2},
\end{cases}
\end{eqnarray}
where $\hat{N}$ denotes the number operator with the property
$\hat{N}|\nu,n\rangle=n|\nu,n\rangle$. The action of the above generators on a set of basis eigenvectors
$|\nu,n\rangle$ is given by
\begin{eqnarray}\label{31}
\begin{cases}
J_0|\nu,n\rangle=(-j+n)|\nu,n\rangle,\\
J_-|\nu,n\rangle=\sqrt{n(-2j+n-1)}|\nu,n-1\rangle,\\
J_+|\nu,n\rangle=\sqrt{(n+1)(-2j+n)}|\nu,n+1\rangle,
\end{cases}
\end{eqnarray}
where $j:=-(\nu+1)/2<0$ denotes the Bargmann index of the dynamical group.
The corresponding Casimir operator can be calculated as
\begin{eqnarray}\label{32}
\begin{cases}
J^2:=J_0(J_0-1)-J_+J_-,\\
J^2|\nu,n\rangle=j(j+1)|\nu,n\rangle,
\end{cases}
\end{eqnarray}
with well known properties
\begin{eqnarray}\label{33}
[J^2,J_{\pm}]=0,\hspace{.3cm}[J^2,J_0]=0.
\end{eqnarray}
Hence, a representation of $su(1,1)$ is determined by the Bargmann index and
the eigenvectors of the Casimir and $J_0$. In addition, we find that the Hamiltonian
(\ref{13}) could be written as \cite{Factorization}
\begin{eqnarray}\label{34}
\begin{array}{cc}
H=72\pi^4\alpha M_{\text{Pl}}\left[J_+J_--(2j+1)J_0\right]\\
\hspace{.5cm}-3\pi^2\alpha M_{\text{Pl}}(j+1)(2j+1)-{\mathcal E},
\end{array}
\end{eqnarray}
from which we can conclude that the Casimir operator (\ref{32}) and $J_0$ commute with
the Hamiltonian
\begin{eqnarray}\label{35}
[J^2,H]=[J_0,H]=0.
\end{eqnarray}
Therefore, $J^2$ and $J_0$ leave the physical Hilbert space invariant and
we choose them as physical operators of the model.
 \section{holographic principle and the minimal area}
 \label{HP}

 Let us first concentrate on a radiation
 dominated very early Universe, $(M=0)$.
 In this case, comparing Eq. (\ref{8}) and (\ref{30m}) gives
 \begin{eqnarray}\label{RR3}
 {\mathcal N}_\gamma=(n+\frac{1}{2})\sqrt{1+36\pi^4\alpha^4}+6\pi^2\alpha^2(n^2+n+\frac{1}{2}).
 \end{eqnarray}
 Moreover, according to Eq.~(\ref{RR1}), ${\mathcal N}_\gamma$ is related
 to the entropy of the radiation fluid. Hence, Eqs. (\ref{RR1}), (\ref{RR3})
 and from the definition of $\nu$, (\ref{28}), lead us extract
 the corresponding entropy of the radiation,
 $S_n^{(\gamma)}$ (in terms of a minimal surface) as
 \begin{eqnarray}\label{40}
 S^{(\gamma)}_n=
 \left(\frac{4\pi^7g}{45}\right)^\frac{1}{4}
 \left(\frac{{\mathcal A}_{\text{min}}}{4G}\right)^\frac{3}{4}(n^2+2n\nu+\nu)^{\frac{3}{4}},
 \end{eqnarray}
 where ${\mathcal A}_{\text{min}}=4\Delta x^2_{\text{min}}=4\alpha^2L_{\text{Pl}}^2$
 is the minimal surface \cite{Nima}. Therefore, according to Eqs. (\ref{31})
 and (\ref{35}), the entropy of radiation is a Dirac observable. To obtain
 a relation between the entropy of radiation and the surface of the apparent
 horizon, let us retrieve the expectation value of the square of the scale factor. Using
  relations (\ref{6}), (\ref{18}), (\ref{23}) and (\ref{30a}),
 we obtain $a(t)=x=i\alpha L_{\text{Pl}}\frac{d}{dz}=\frac{i\alpha
 L_\text{Pl}}{2}(TA-A^\dagger T^\dagger).$
Hence, the expectation value of the square of the scale factor reads
\begin{eqnarray}\label{ex}
\begin{array}{cc}
\langle a^2\rangle=\langle x^2\rangle=\frac{\alpha^2L_{\text{Pl}}^2}{4}\langle\nu,n|\left(TAA^\dagger T^\dagger+A^\dagger A\right)|\nu,n\rangle\\
=\frac{{\mathcal A}_{\text{min}}}{8}(n^2+2n\nu+\nu+n-\frac{1}{2}),
\end{array}
\end{eqnarray}
where we have used (\ref{30b}).
In the presence of the minimum length, from relation~(\ref{28}), we get $\nu\simeq1$. Hence for large
values of the quantum number, $n$, we find $\langle a^2\rangle\simeq{\mathcal
A}_{\text{min}}n^2/8$. On the other hand, the apparent horizon of
a FLRW model for the radiation case is given~\cite{F11} by
$R_{\text{ah}}=(H^2+1/a^2)^{-1/2}=\sqrt{\frac{6\pi^2M^2_{\text{Pl}}}{\mathcal
N}}a^2$.
Inserting this result into Eq. (\ref{40}), we find, for large values of $n$, that
\begin{eqnarray}\label{hooft}
S^{(\gamma)}_n\simeq\Big(\frac{2048\pi^7g}{45}\Big)^\frac{1}{4}\left(\frac{{\mathcal
A}_{\text{ah}}}{4G}\right)^{\frac{3}{4}},
\end{eqnarray}
where ${\mathcal A}_{\text{ah}}:=4\pi\langle R_{\text{ah}}\rangle^2$ denotes
the area of the apparent horizon. The above equation is in the form as
conjectured by 't Hooft \cite{Hooft}: assuming that
the matter occupies a specific volume, then
the entropy of that matter will be $S^{(\gamma)}\propto
(4G)^{-3/4}{\mathcal A}^{3/4}$ \cite{Hooft},
where ${\mathcal A}$ denotes the area of the containing volume.

We now turn to a Universe filled with only dust fluid, $({\mathcal N}_\gamma=0)$.
In this case, comparing Eqs. (\ref{8}) and (\ref{30m}), we get
 \begin{eqnarray}\label{PP1}
 M^2=144\pi^4\alpha^2M^2_{\text{Pl}}(n^2+2n\nu+\nu).
 \end{eqnarray}
In addition, let us discuss the total entropy of the dust content of the Universe,
by means of investigating the following expression~\cite{Kittel} for the entropy of an ideal
gas, which consists of $N$ ideal particles in a volume $V$, namely,
\begin{eqnarray}\label{BB1}
S_{\text{(ideal)}}=N\ln\left(\frac{V}{N}\left(\frac{mT}{2\pi}\right)^{\frac{3}{2}}e^\frac{5}{2}\right),
\end{eqnarray}
 where $m$ is the mass of the particles. For the case
of a continuous fluid, let us rewrite~(\ref{BB1}).
To this end, we consider an
ideal gas contained within a small volume element $dV$.
The number of particles inside $dV$ is
\begin{eqnarray}\label{BBB2}
dN=\frac{\rho}{m}dV.
\end{eqnarray}
 Inserting this expression, into
Eq.~(\ref{BB1}), the entropy associated with the
volume element, in terms of the density of the fluid, can be written as
\begin{eqnarray}\label{BB3}
dS^{\text{(dust)}}=\frac{\rho_m}{m}\ln\left(\frac{KT^\frac{3}{2}}{\rho}\right)dV,
\end{eqnarray}
where $K=(\frac{m^5e^5}{2\pi})^{1/2}$ \cite{Gron}.
For a dust dominated Universe where $\rho=\rho_0(a/a_0)^{-3}$ and
$T=T_0(a/a_0)^{-2}$, we get $S^{\text{(dust)}}=\ln(KT_0^{3/2}/\rho_0)N$.
Let us use the
simple approximation
\begin{eqnarray}\label{BB4}
S^{\text{(dust)}}\simeq N=\frac{M}{m},
\end{eqnarray}
Therefore, from Eqs. (\ref{PP1}) and (\ref{BB4}), we obtain
\begin{eqnarray}\label{BB5}
S_n^{\text{(dust)}}\simeq 12\pi^2\alpha\frac{M_{\text{Pl}}}{m}\sqrt{n^2+2n\nu+\nu}.
\end{eqnarray}
 For this dust Universe, the apparent horizon~\cite{F11}
is $R^2_{\text{ah}}=\frac{6\pi^2M_{\text{Pl}}^2}{M}a^{3}$.
Employing the relation associated to the variable $x$ from (\ref{6}) and
reminding that the expectation values of odd powers
of $x$ vanish, i.e., $\langle x\rangle=\langle x^3\rangle=0$,
we obtain $\langle a^3\rangle=-x_0[3\langle x^2\rangle+x_0^2]$.
Now, substituting the relation for $x_0$, which is defined as in (\ref{6}),
gives $\langle a^3\rangle=\frac{M}{12\pi^2M_{\text{Pl}}^2}[3\langle x^2\rangle+x_0^2]$.
The terms inside the brackets can be replaced by the
expectation value of $x^2$ from (\ref{ex}) and an expression for $x_0^2$ as
\begin{eqnarray}\label{AB1}
x_0^2=\frac{{\mathcal A}_{\text{min}}}{4}(n^2+2n\nu+\nu),
\end{eqnarray}
where we have used (\ref{6}), $\alpha^2=({\mathcal A}_{\text{min}}M_{\text{Pl}}^2)/4$
and the quantization condition (\ref{PP1}).
Consequently, using Eqs. (\ref{ex}) and (\ref{AB1}) in
the last expression associated to the $\langle a^3\rangle$, and finally
substituting it into the relation of the apparent horizon, we find
\begin{eqnarray}\label{42s}
\begin{array}{cc}
\langle R_{\text{ah}}^2\rangle=\frac{1}{2}\left[3\langle x^2\rangle+x_0^2\right]=\\
\frac{{\mathcal A}_{\text{min}}}{16}[5(n^2+2n\nu+\nu)+3(n-\frac{1}{2})].
\end{array}
\end{eqnarray}
Therefore, at the presence of minimal length and for large values of the
quantum number $n$, by comparing (\ref{BB5}) and (\ref{42s}) we obtain
\begin{eqnarray}\label{BB6}
S^{(\text{dust})}=\frac{12\pi}{\sqrt{10}}\frac{M_{\text{Pl}}}{m}\left(\frac{{\mathcal
A}_{\text{ah}}}{4G}\right)^\frac{1}{2},
\end{eqnarray}
 which implies that the t' Hooft holographic conjecture
is satisfied for a Universe filled with a dust fluid.
Let us, in particular, consider the specific case
where primordial black holes (PBHs)
constitute the sole content of the dust fluid~\cite{B7}.
 In this case, Eq. (\ref{PP1})
gives us the total entropy of the PBHs: If we assume each PBH to have the same mass, $m_{\text{bh}}$,
then we have $M=Zm_{\text{bh}}$, where $Z$ denotes the total number of PBHs.
 The area of the event horizon of a Schwarzschild black hole is given by ${\mathcal A}_{\text{eh}}=16\pi
 G^2m_{\text{bh}}^2$ and consequently we obtain $M^2=4\pi Z^2{\mathcal A}_{\text{eh}}M_{\text{Pl}}^4$,
   assuming each PBH to be of a Schwarzschild type.
 Therefore, from Eq. (\ref{PP1}) we can obtain the following relations
 between the PBH event horizon area and the minimal surface area
 \begin{eqnarray}\label{41}
  {\mathcal A}_{\text{eh}}\simeq\frac{9\pi^3}{Z^2}{\mathcal A}_{\text{min}}(n^2+2n\nu+\nu).
 \end{eqnarray}
  Hence, the event horizon of PBHs are Dirac observables, with the event
 horizon being related to
 the quantum number $n$.
 Using the Bekenstein-Hawking formula $S^{(\text{bh})}=\frac{{\mathcal A}_{\text{eh}}}{4G}$,
 we obtain the total entropy of the PBHs, $(ZS^{(bh)})$, as
\begin{eqnarray}\label{42}
S^{\text{(PBHs)}}_n=
\frac{9\pi^3}{Z}\frac{{\mathcal A}_{\text{min}}}{4G}(n^2+2n\nu+\nu).
\end{eqnarray}
Therefore, for large values of quantum number $n$, using (\ref{42s}) we obtain
\begin{eqnarray}\label{43}
S^{\text{(PBHs)}}=\frac{81\pi^3}{7Z}\left(\frac{{\mathcal A}_{\text{ah}}}{4G}\right),
\end{eqnarray}
which is in agreement with t' Hooft holographic conjecture \cite{Hooft}.
\section{conclusions}
\label{con}
In this paper we studied the effects of a deformed Heisenberg
algebra in terms of a MLUR in a closed quantum FRLW model, whose matter
is either a fluid of radiation or dust.
 Quantum cosmologies with
a perfect fluid matter content were investigated, e.g., in \cite{Fluid1}.
In particular, the case of a dust or radiation dominated quantum Universe
was studied in \cite{Fluid2}.

Our main result is that the extended dynamical group of the model, $su(1,1)$, admits a minimal
area retrieved from a MLUR, in the form of a Dirac observable.
It reasonably agrees with the cosmological holographic
principle, in the case of a large quantum number, $n$.
We are aware that our results are obtained within a very simple as well as restricted setting.
Nevertheless, we think they are intriguing and provide motivation for subsequent research works.
Possible extensions to test the relation among a GUP, a minimal surface being subsequently
obtained (constituting a Dirac observable) and the cosmological holography, may include:\\
  (i) Considering other perfect fluids besides radiation and dust.\\
  (ii) Including instead, e.g., scalar fields.\\
  (iii) Considering a Bianchi IX geometry.\\
  (iv) Exploring string features by means of a broader gravitational sector.\\
\section{ACKNOWLEDGMENTS}
The authors thank H. R. Sepangi for reading the manuscript.
S. M. M. Rasouli is grateful for the support of grant SFRH/BPD/82479/2011
from the Portuguese Agency Funda\c{c}\~{a}o para a Ci\^{e}ncia e Tecnologia.
This research work was supported by the grant
CERN/FP/123618/2011 and PEst-OE/MAT/UI0212/2014.


\begin{thebibliography}{99}
\bibitem{footnote}These expressions can be used, with
 $g=g_b+\frac{7}{8}g_f$, where $g_b$ and $g_f$ are the total number
 of internal degrees of freedom of all
 ultra-relativistic bosons and fermions, respectively.
\end{thebibliography}
\end{document}